\documentclass[a4paper,11pt]{article}
\usepackage{pos}
\usepackage{units}
\usepackage{subcaption}
\usepackage{natbib}
\newcommand{\Offline}{\mbox{$\overline{\rm Off}$\hspace{.05em}\raisebox{.3ex}{$\underline{\rm line}\enspace$}}}

\title{\Offline simulation and reconstruction software framework for the JEM-EUSO missions}
 \ShortTitle{JEM-EUSO \Offline software framework}

\author*[a]{T. Paul}

\affiliation[a]{Dept. of Physics, Lehman College, City University of New York\\
  250 Bedford Park Boulevard West, Bronx, New York, USA \\
}

\onbehalf{for the JEM-EUSO Collaboration\\[-1mm]{\normalsize \normalfont (a complete list of authors can be found at the end of the proceedings)}}

\emailAdd{Thomas.Paul@lehman.cuny.edu}

\abstract{ The Joint Exploratory Missions for an Extreme Universe Observatory
  comprises a collection of complementary missions dedicated to pioneering
  technologies and techniques for a future space-based multi-messenger
  observatory which will have sufficient sensitivity and exposure to measure
  properties of extremely rare ultra-high energy (E>50 EeV) cosmic rays and very
  high energy (E>100 PeV) neutrinos.  Here we describe a general-purpose
  software framework designed to facilitate detailed simulation and
  reconstruction of events observed by the various missions using both detection
  of fluorescence and Cherenkov light produced when cosmic ray or neutrino
  induced extensive air showers traverse Earth's atmosphere. The software builds
  on a framework developed by the Pierre Auger Collaboration. We describe the
  techniques used to organize contributions from numerous collaborators, manage
  an abundance of configuration information, and provide simple access to
  time-dependent detector and atmospheric information. We also explain how we
  seamlessly support a multitude of computing platforms, provide fast
  installation and maintain the broad testing coverage required for stability of
  the large and heterogeneous code base. We provide a few examples of simulated
  and reconstructed data gathered by some of the JEM-EUSO missions, including
  the EUSO-SPB2 instrument.  }

\ConferenceLogo{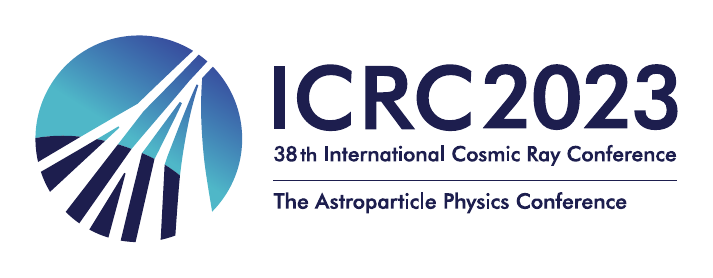}

\FullConference{%
38th International Cosmic Ray Conference (ICRC2023)\\
  26 July - 3 August, 2023\\
  Nagoya, Japan}

\begin{document}
\maketitle

\section{Introduction}
\label{sec:intro}

One challenge confronting large, geographically dispersed collaborations is how
to effectively manage the development of software for simulation and
reconstruction of data in a way that allows contributions from many
collaborators to be straightforwardly accommodated and maintained. The Joint Exploratory
Missions for Extreme Universe Observatory (JEM-EUSO)
Collaboration~\cite{JEM-EUSO} comprises hundreds of scientists from 10
countries, and thus faces the same challenges as other ``big science''
collaborations.

The JEM-EUSO~\cite{JEM-EUSO} missions consist of complementary campaigns
in pursuit of a future space-based multi-messenger observatory, such as
POEMMA~\cite{POEMMA}, with the ultimate objective of uncovering the
origins and composition of Ultra-High-Energy ($E > 20~\rm{EeV}$)
Cosmic Rays (UHECR), and discovery of very high energy ($E >
20~\rm{PeV}$) neutrinos originating from astrophysical transient
sources~\cite{transients}.

Several pathfinder missions have been developed to demonstrate the
technologies required to achieve this objective. A first prototype
instrument was carried aboard a one-night high-altitude balloon
flight in 2014~\cite{EUSO-Balloon}. This instrument employed
Fresnel optics and one photo-detector module (PDM) and was underflown by helicopter-borne light sources to test the technology.  In 2017, a long-duration
super pressure balloon flight was launched employing the same basic detection
technique with the goal of detecting fluorescence
light emitted when cosmic ray air showers excite atmospheric nitrogen
as they traverse the Earth's atmosphere. Only limited data were gathered due to an apparent balloon flaw~\cite{EUSO-SPB1}.
The EUSO-TA instrument, which also comprises 1 PDM and Fresnel optics, is
deployed adjacent to a Telescope Array (TA) fluorescence station~\cite{EUSO-TA}
and can detect both air showers and laser shots. Mini-EUSO, a scaled-down
instrument, has been taking data aboard the International Space Station (ISS)
since 2019~\cite{Mini-EUSO}. A second long-duration super pressure balloon
flight, EUSO-SPB2~\cite{EUSO-SPB2}, was launched on May 13 2023. This mission carried two
Schmidt telescopes, one pointed downward to detect fluorescence light and
one with adjustable pointing in the vicinity of the Earth's limb to detect Cherenkov light from
nearly-horizontal showers and to search from up-going showers (below the limb) produced
by astrophysical transients, and to study backgrounds. Unfortunately, this
balloon was also flawed, and despite good detector performance, limited data were collected.

From the early stages of JEM-EUSO planning, a software package called EUSO
Simulation and Analysis Framework (ESAF)~\cite{esaf} was implemented in order to
address large-scale simulation and reconstruction challenges of an instrument
originally proposed to fly aboard the ISS~\cite{jemeuso}. At roughly the same
time, the Pierre Auger Collaboration~\cite{ThePierreAuger:2015rma} was
developing software to address similar challenges, with attention to building in
the flexibility to accommodate future (unspecified-at-the-time) extensions. The
design of the Auger \Offline software~\cite{Argiro:2007qg} proved successful in
this regard and has been extended for upgrade requirements~\cite{extensions} and
adopted in part by the NA61/SHINE Collaboration~\cite{shine}. It thus seemed
natural to adopt portions of the Auger \Offline software for the requirements of
JEM-EUSO, including the overarching framework and utilities, as well as
algorithms related to fluorescence and Cherenkov light emission and propagation
that have been vetted with Auger Observatory data.

Here we briefly review the \Offline design and describe the extensions that have
been developed for the various JEM-EUSO pathfinder missions. Further details
are available in~\cite{Argiro:2007qg, newoffline}.

\section{Framework overview}
\label{sec:framework}
The overall Auger \Offline framework has been largely retained for
EUSO-\Offline, and is outlined here both for completeness and to illustrate
some of the ways the flexibility of the software facilitates
simulation and reconstruction for a variety of pathfinder instruments.
More details can be found in~\cite{Argiro:2007qg, newoffline}.

\begin{figure}[ht]
\centering
\includegraphics[width=0.35\textwidth]{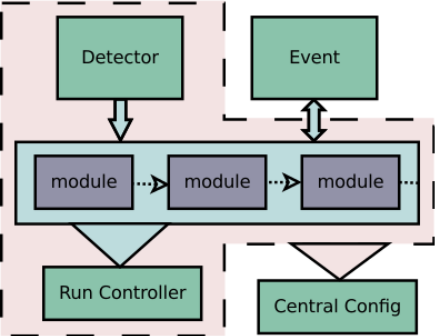}
\caption{General organization of the \Offline framework. See the text for details.}
\label{f:modules}
\end{figure}

The framework comprises physics algorithms contained in {\em Modules}; a {\em RunController}
which commands module execution; a read/write {\em Event Data Model} (EDM) from which modules read
information and to which they write their results; a {\em Detector Description} (DD)
which provides an interface to look up detector properties and conditions data; and a {\em CentralConfig} which
points the modules and framework components to the configuration data they require, and
which tracks provenance in order to support repeatability. The general scheme is illustrated in
Fig.~\ref{f:modules}.

Simulation and reconstruction applications can generally be organized
as a sequential pipeline of algorithms, with each encapsulated in a
{\em module} which is registered with the {\em RunController}. The
common module interface enables simple swapping out of algorithms
in order to compare different approaches to a given problem, or to
build up different applications. For instance, one can choose a module
to generate a simulated signal from different sources, such as simulated
air showers, laser shots, or other light sources, without having
to recompile any code. Steering the execution of the modules is
performed using a simple XML file which is read by the {\em RunController}.

The EDM itself is configurable (using XML) so that it can reflect the data
structures of different JEM-EUSO missions.  It is also equipped with a
mechanism allowing modules to check the EDM constituents in order to determine if
they can apply their algorithms to the event, or if some other action is required.

Data related to the instrument and time-dependent conditions data are accessible
via the DD interface. The DD relays requests for information to a configurable
back-end, known as the {\em Manager}, which can retrieve the requested data from
different sources, such as databases or XML files. In this way, the DD interface
code remains relatively simple, while the back-end handles the ``dirty work'' of
finding and reading data in different formats.  The Atmosphere and Earth are a
part of the DD.  A plug-in mechanism in the atmosphere description supports
various interchangeable {\em models} for computing fluorescence and Cherenkov
yields as well as Rayleigh and Mie scattering and Ozone absorption. Models can
use either parametrizations written in XML files or measurements stored in databases. The Earth
provides access to albedo estimates of different terrain, which can be
used to simulate Cherenkov light reflected from the Earth surface into an
orbiting telescope.

The \Offline framework provides an XML and XML Schema-based configuration
system with sufficient flexibility to accommodate the different JEM-EUSO instruments.
This system is also employed to organize parameters
used in individual physics modules. A {\em Central Config} (CC) object directs
modules and framework components to the data they require. 
The CC also records all configuration data used during a run and stores it
in an XML file which can be read back in order to reproduce a given run.

\section{DevOps}
\label{sec:devops}

A significant effort has been devoted to writing thorough tests, which are
run automatically with the help of the Continuous
Integration / Continuous Deployment (CI/CD) system provided by GitLab. This is
particularly crucial for a project like EUSO-\Offline, which is designed to
support several past, ongoing, and future missions; new instrument designs have
to be accommodated without breaking the old ones. We test the low-level code
with a battery of unit tests built with the help of CppUnit (for older tests) and
GoogleTest (for newer ones). Regression tests are performed on full simulation
and reconstruction applications using a set of in-house tools for serializing
results to check against references. A set of Python tools has been developed to
find and run regression tests and example applications in parallel (using Dask)
on the CI/CD for quick turnaround.
We also exploit linting and static analysis tools provided by the Clang project.

EUSO-\Offline is built using CMake to write the build tool (either GNU Make or
ninja). External dependencies include ROOT, Xerces, Geant4, boost, mysql,
pytorch-c as well as a number of Python packages. We use Anaconda (a.k.a. conda)
together with Libmamba to solve all of the dependencies and install the Python
packages as well as pre-compiled binaries for the C/C++
dependencies. Installation via an industry standard like conda/mamba is robust
and portable and can be performed in a few minutes rather than the many hours
typically required to compile large packages like ROOT and Geant4 from source.

\section{Simulation and Reconstruction in EUSO-\Offline}

One of the benefits of inheriting portions of the Auger \Offline code
is that it provides a set of well-tested modules for simulating
light generation and propagation for both air showers and laser shots,
various models for computing fluorescence yield, including the latest
experimental data, as well as models for computing Rayleigh and Mie
scattering and absorption using either parametric models or time-dependent
measurements stored in a database. For the JEM-EUSO missions, it was
necessary to extend the code in a number of ways, itemized below.

\begin{itemize}
\item The effects of Ozone absorption have been included as a {\em model}
  accessible via the Atmosphere interface.
  \vspace{-0.2cm}
\item The Earth interface has been added to the DD to provide access to
  albedo estimates of different sorts of terrain which can be used
  for simulation of Cherenkov light reflected from the earth's surface
  into an orbiting telescope.
  \vspace{-0.2cm}
\item A reader for the EASCherSim~\cite{easchersim} Monte Carlo generator has been written.
  This generator supports the simulation of Cherenkov light emitted at very
  small angles with respect to the shower axis.
  \vspace{-0.2cm}
\item A very configurable Geant4-based telescope simulation module
  has been written, which can model any of the JEM-EUSO instrument designs
  by simply specifying the desired XML configuration.
	\vspace{-0.2cm}
\item Custom Fresnel optics simulation (written originally for ESAF) has
  been incorporated into the Geant4 simulation of the telescopes.
	\vspace{-0.2cm}
\item Background simulation modules (both night-glow and spot-like) have been
  prepared and can model background light observed by the Fluorescence
  telescopes in either tilt or nadir pointing modes.
	\vspace{-0.2cm}
\item Simulation of the trigger logic for the different instruments has been
  prepared.
  \vspace{-0.2cm}
\item  A convolutional neural network was developed and trained on simulated data
  to perform fast in-flight classification of events in order to prioritize
  event downlinking~\cite{george}.
\end{itemize}

\begin{figure}[htb]
 	\centering
 	\begin{subfigure}[a]{0.59\textwidth}
     	\centering
     	\includegraphics[width=\textwidth]{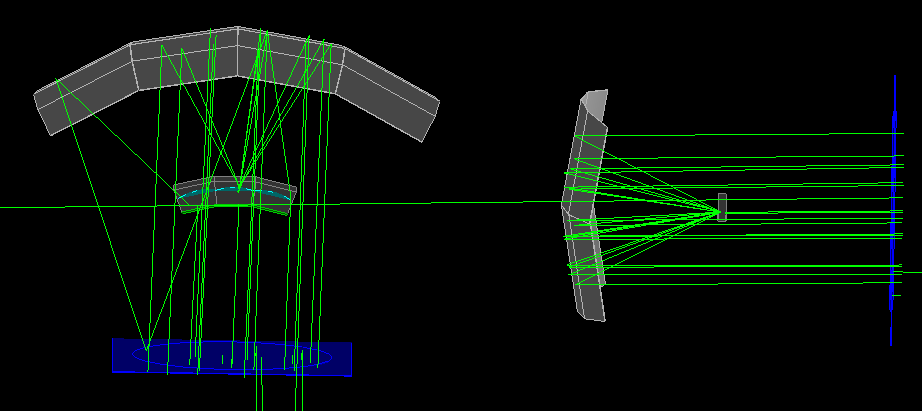}
     	\caption{
        EUSO-SPB2 instrument. The green lines
       	represent photons. Mirrors and cameras are
       	shown in grey, and the blue region represents the entrance pupil.
       	The fluorescence telescope is on the right and the Cherenkov telescope is on the left.
     	}
 	\end{subfigure}
 	\hfill
 	\begin{subfigure}[a]{0.39\textwidth}
     	\includegraphics[width=\textwidth]{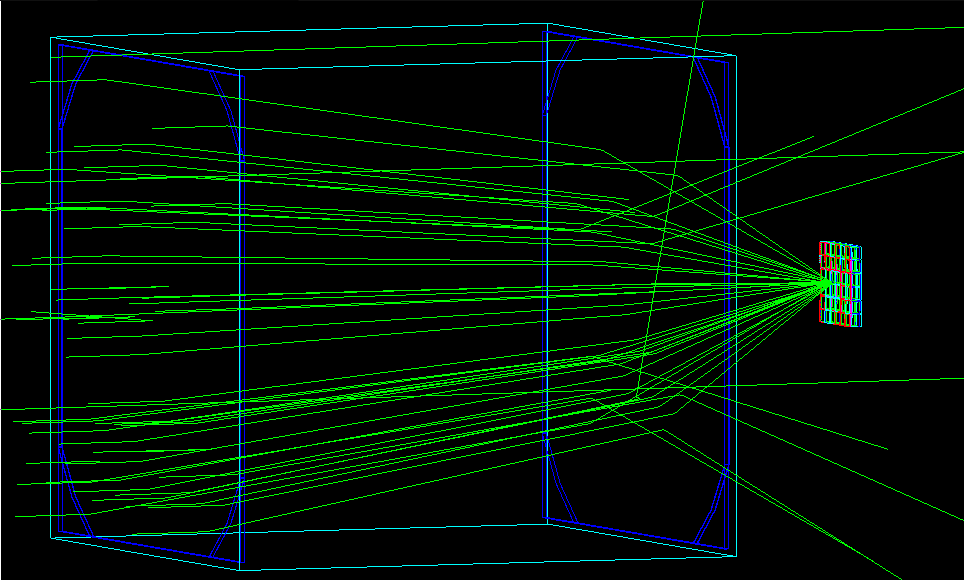}
     	\caption{
       	EUSO-TA instrument, comprising 2 Fresnel lenses and 1 PDM.
     	}
 	\end{subfigure}
    \caption{Geant4 simulations of SPB2 and EUSO-TA telescopes.}
	\label{f:geant4}
\end{figure}
We now broadly outline how various modules are assembled to perform simulation
and reconstruction of data.
Simulations begin with a module that reads output from a cosmic ray air shower generator or a
simulation of a laser shot. A common interface, described in
~\cite{Argiro:2007qg}, connects the \Offline to all the generator readers.
A subsequent module positions the shower relative to the telescope. Cuts are
applied based on the telescope's Field of View (FoV) in order to increase
simulation speed. Subsequent modules compute the fluorescence and Cherenkov
emission and propagation, including Rayleigh and Mie scattering and Ozone
absorption based on models accessible via the Atmosphere
interface, which follow parameterizations and data from~\cite{models}.
The subsequent Geant4-based Simulator performs ray-tracing of the photons
through the telescope optical system, as illustrated in Fig.~\ref{f:geant4}
for the cases of EUSO-SPB2 and EUSO-TA.

\begin{figure}[htb]
 	\centering
 	\begin{subfigure}[a]{0.4\textwidth}
     	\centering
     	\includegraphics[width=\textwidth]{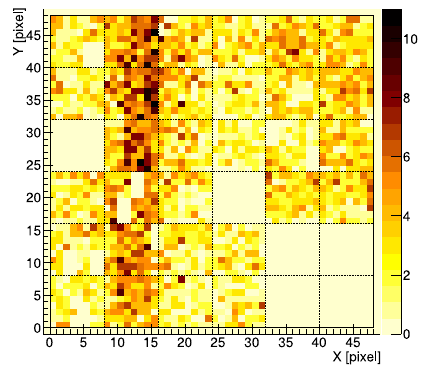}
     	\caption{Data}
 	\end{subfigure}
 	\hfill
 	\begin{subfigure}[a]{0.4\textwidth}
     	\centering
     	\includegraphics[width=\textwidth]{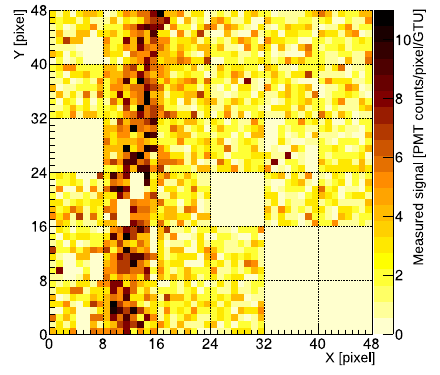}
     	\caption{Simulation}
 	\end{subfigure}
	\caption{A UHECR track with an energy of
  	\unit[10$^{18}$]{eV} and an impact point \unit[2.6]{km} from the
  	detector, according to TA reconstruction. The zenith angle of the shower was 8\textdegree~ and its azimuth
  	was 82\textdegree. Panel (a) shows the track as recorded in EUSO-TA while
  	panel (b) shows the \Offline simulation of a shower with the parameters
  	from TA. Figure from \cite{EUSO-TA}.}
	\label{fig:datasim}
\end{figure}

A sequence of downstream modules simulates the camera efficiency,
electronics response, and digitization based on laboratory measurements.
Further modules simulate the backgrounds (described above) and
the trigger logic. At this stage, the fully simulated shower can be written
to file or passed directly to the sequence of reconstruction
modules. Figure~\ref{fig:datasim} shows a shower recorded by
EUSO-TA next to a shower simulated using the parameters
taken from the TA measurement.
\begin{figure} [ht]
  \centering
  \includegraphics[width=0.6\textwidth]{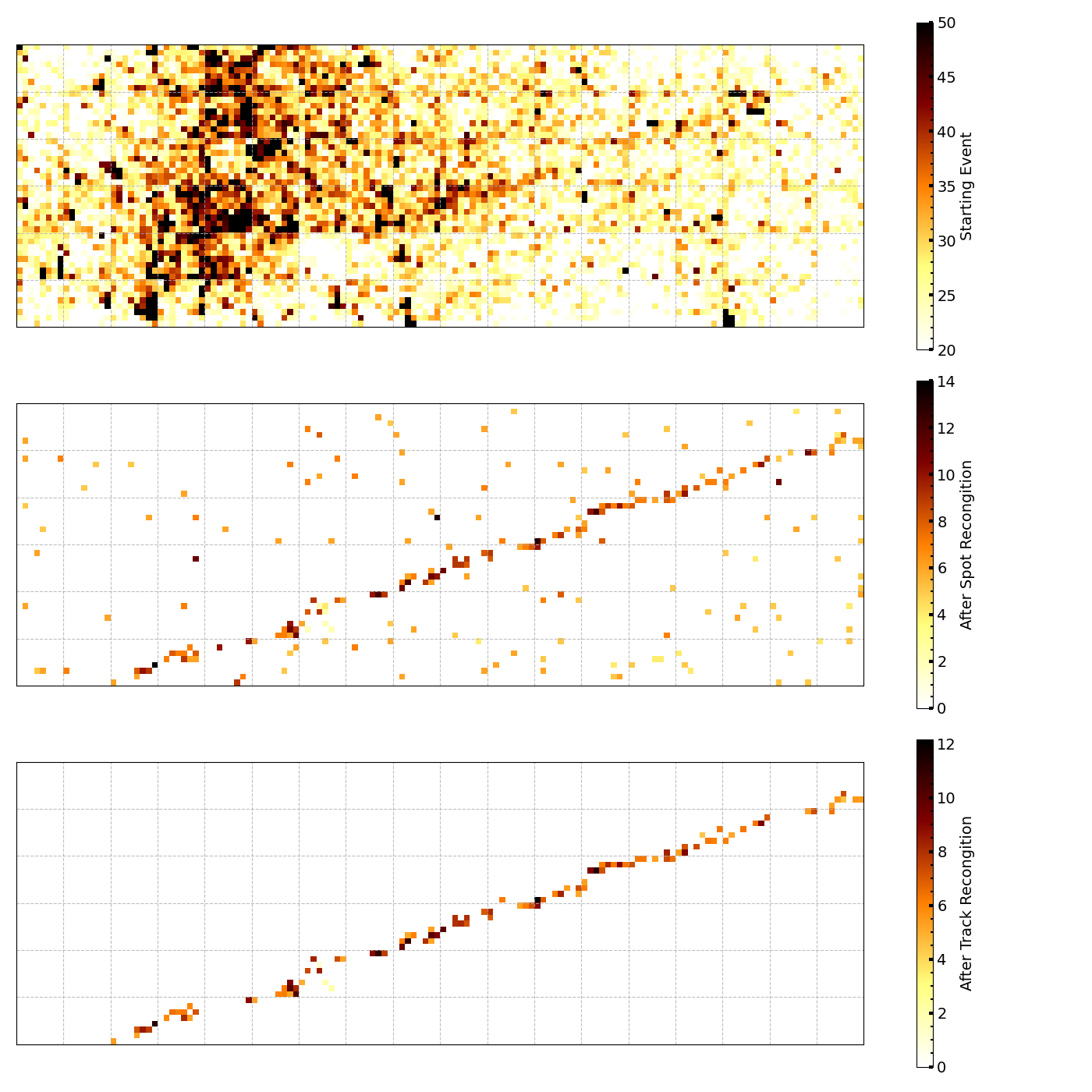}
  \caption{Laser track recorded on the EUSO-SPB2 fluorescence detector focal surface during the field
	campaign in the summer of 2022 at the TA site.
	Integrated counts over 128 time frames (top), pixels remaining after spot identification (middle), and pixels remaining after track identification (bottom).}
\label{fig:BackgroundSubtraction}
\end{figure}
\vspace{-0.2cm}
Like simulation, reconstruction is performed by a Module sequence, which
ultimately produces an estimate of the primary cosmic ray's energy, arrival
direction, and composition. Backgrounds first are removed from the data
pixel-by-pixel with a threshold cut based on the mean and standard deviation of
the pixel trace. Subsequent modules further clean the track using
constraints on pixel isolation from tracks and clusters in both space and time.
An example of this track cleaning for the EUSO-SPB2 fluorescence telescope
is shown in Fig.~\ref{fig:BackgroundSubtraction}.

The shower-detector plane is computed from the pointing and signal size of
pixels in the track. Using the shower-detector plane, the elevation angle of
each pixel is determined, and a fit to the time-elevation distribution is used
to extract the shower's arrival direction (see eg. ~\cite{jem-euso:2018tmw}).
A final module uses a fit to a Gaisser Hillas function to estimate the
shower energy and atmospheric depth where the shower maximum occurs ($X_{\rm max}$).

\vspace{-0.2cm}
\section{Summary}
Adapting the Auger \Offline framework to the JEM-EUSO program has been a
relatively straightforward process owing to the modular design, the flexibility
provided by the configuration system, and the collection of well-vetted modules to
assist in simulations. We have extended the system, predominantly with new
modules specific to the JEM-EUSO mission characteristics, while modernizing the
code base with newer tools for testing, building and deployment, and adoption
of modern C++ standards.

\acknowledgments
\vspace{-.3cm}
{\footnotesize
The authors would like to thank the Pierre Auger Collaboration for providing the framework code as well as portions of the simulation code.
The authors would like to acknowledge the support by NASA award 11-APRA-0058, 16-APROBES16-
0023, 17-APRA17-0066, NNX17AJ82G, NNX13AH54G, 80NSSC18K0246, 80NSSC18K0473,
80NSSC19K0626, 80NSSC18K0464, 80NSSC22K1488, 80NSSC19K0627 and 80NSSC22K0426.
This research used resources of the National Energy Research Scientific Computing Center (NERSC),
a U.S. Department of Energy Office of Science User Facility operated under Contract No. DE-AC02-
05CH11231. We acknowledge the NASA Balloon Program Office and the Columbia Scientific Balloon Facility and staff for extensive support.
We acknowledge the ASI-INFN agreement n. 2021-8-HH.0 and its amendments.
We also acknowledge the invaluable contributions of the administrative and technical staffs at our home institutions.
}
\vspace{-0.4cm}
\setlength{\bibsep}{6pt}

%
%
%
\newpage
{\Large\bf Full Authors list: The JEM-EUSO Collaboration\\}

\begin{sloppypar}
{\small \noindent
S.~Abe$^{ff}$, 
J.H.~Adams Jr.$^{ld}$, 
D.~Allard$^{cb}$,
P.~Alldredge$^{ld}$,
R.~Aloisio$^{ep}$,
L.~Anchordoqui$^{le}$,
A.~Anzalone$^{ed,eh}$, 
E.~Arnone$^{ek,el}$,
M.~Bagheri$^{lh}$,
B.~Baret$^{cb}$
% ...
}
\end{sloppypar}
\vspace*{.3cm}

{ \footnotesize
\noindent
% ...
$^{cb}$ Universit\'e de Paris, CNRS, AstroParticule et Cosmologie, F-75013 Paris, France\\
% ...
$^{ed}$ Istituto Nazionale di Fisica Nucleare - Sezione di Catania, Italy\\
% ...
$^{eh}$ INAF - Istituto di Astrofisica Spaziale e Fisica Cosmica di Palermo, Italy\\
% ...
$^{ek}$ Istituto Nazionale di Fisica Nucleare - Sezione di Torino, Italy\\
$^{el}$ Dipartimento di Fisica, Universit\`a di Torino, Italy\\
% ...
$^{ep}$ Gran Sasso Science Institute, L'Aquila, Italy\\
% ...
$^{ff}$ Nihon University Chiyoda, Tokyo, Japan\\ 
% ...
$^{ld}$ University of Alabama in Huntsville, Huntsville, AL, USA\\
$^{le}$ Lehman College, City University of New York (CUNY), NY, USA\\
% ...
$^{lh}$ Georgia Institute of Technology, USA\\
% ...
}

\end{document}